\documentstyle[aps]{revtex}
\begin{document}
\draft
\title{Energy-Momentum Tensor Valued Distributions for the Schwarzschild and
Reissner-Nordstrom Geometries}
\author{N. R. Pantoja\footnotemark[1]  and H. Rago\footnotemark[2] }
\address{\footnotemark[1]  Centro de Astrof\'{\i}sica Te\'orica. \\
Departamento de F\'{\i}sica. Facultad de Ciencias.\\
Universidad de los Andes. M\' erida 5101. Venezuela.\\
\footnotemark[2]  Laboratorio de F\'{\i}sica Te\'orica and \\
Centro de Astrof\'{\i}sica Te\'orica. \\
Departamento de F\'{\i}sica. Facultad de Ciencias.\\
Universidad de los Andes. M\' erida 5101. Venezuela.}
\maketitle

\footnotetext[1]{pantoja@ciens.ula.ve} 
\footnotetext[2]{rago@ciens.ula.ve}

\begin{abstract}
An approach to computing, withing the framework of distribution theory, the
distributional valued energy-momentum tensor for the Schwarzschild spacetime
is disscused. This approach avoids the problems associated with the
regularization of singularities in the curvature tensors and shares common
features with the by now standard treatment of discontinuities in General
Relativity. Finally, the Reissner-Nordstrom spacetime is also considered
using the same approach.
\end{abstract}

\pacs{PACS numbers: 0420, 0470B, 0250N, 0420C,9760L}

\section{Introduction}

The energy-momentum tensor ${\bf T}$ associated to a point particle has an
unambiguos character either if the particle is considered in a flat
minkowskian spacetime or as a test particle in a curve background geometry.
Obtained by a constructive procedure, invoking the minimal coupling
principle or defined as the variational derivative of the point particle
action with respect to the metric, the result in an obvious notation, is 
\begin{equation}
T_{\ }^{ab}=m\frac{\delta (\vec x-\vec x_p)}{\sqrt{-g}}u^a\frac{dx^b}{dt}
\label{intro1}
\end{equation}
If the particle is at rest with respect to the chosen reference frame, the
only surviving component is the $t-\!t$ component, i.e., the one
representing energy density, as it should be for a particle with zero
momentum. 

The situation changes radically in general relativity if the point particle
is acting as the gravitational field source. The singular behaviour of the
metric and related geometrical quantities at the particle location turns the
mathematical meaning of the energy tensor rather imprecise. Nevertheless, on
physical grounds or suggested by application of the variational principle
leading to Einstein field equations, it seems plausible that the only non
vanishing component of the energy-momentum tensor for a static point source
is again the $t-\!t$ component, with support at the particle position, and
that is what it should be expected, whatever techniques are employed to find 
${\bf T}$. 

In the classical theory of gravitation one is led to consider the Einstein
field equations which are, in general, quasilinear partial differential
equations involving second order derivatives for the metric tensor. Hence,
continuity of the first fundamental form is expected and at most,
discontinuities in the second fundamental form, the coordinate independent
statements appropriate to consider 3-surfaces of discontinuity in the
spacetime manifolfd of General Relativity \cite{israel}.

The use of classical distribution theory, in order to incorporate singular
information into General Relativity, was advocated some time ago \cite
{parker,raju}. However, it has been shown that the singular parts of the
Riemann tensor must be of support on a submanifold of dimension $d \geq 3$,
in order to have well defined tensor distributions for it and its
contractions \cite{geroch}. In spite of this, several attemps have been made
to regularize the Schwarzschild \cite{balasin93,kawai} and Kerr-Newman
spacetimes \cite{balasin94,balasin95} to obtain, after suitable limiting
processes, the distributional energy-momentum tensors which play the role of
point sources for these geometries with results that seem to depend on the
regularization prescription \cite{kawai}.

Besides the fact that the notion of a point source may has no precise
mathematical meaning withing the framework of the nonlinear Einstein's
theory of general relativity, some failures of previous works can be, to
some extend, clearly stablished and then circumvented. In order to reveal
some of the weaknesses of the above forementioned regularization approaches,
we will rederive the results of Ref.\cite{balasin93} following a different
approach. Consider the line element 
\begin{equation}
ds^2=h(r)dt^2-h(r)^{-1}dr^2+r^2d{\theta }^2+r^2sin^2\theta \,\,d{\phi }^2,
\label{schcurv}
\end{equation}
where 
\begin{equation}
h(r)=-1+\frac{2m}r\Theta (r-\epsilon ),  \label{curvreg}
\end{equation}
with $r<2m$, $\Theta (r-\epsilon )$ is the Heaviside function and the limit $%
\epsilon \rightarrow 0$ is understood. Equation (\ref{schcurv}) with $h$ as
given in (\ref{curvreg}) can be considered as a regularized version of the
Schwarzschild line element in curvature coordinates. From equation (\ref
{schcurv}), the calculation of the Einstein tensor proceeds in a
straighforward manner. It gives 
\begin{equation}
G_{\,\,t}^t=G_{\,\,r}^r=-\frac 1r\frac{dh}{dr}-\frac 1{r^2}(1+h),
\end{equation}
and 
\begin{equation}
G_{\,\,\theta }^\theta =G_{\,\,\phi }^\phi =-\frac 12\frac{d^2h}{dr^2}-\frac 
1r\frac{dh}{dr}.
\end{equation}
Next, taking the derivatives in the sense of distributions one finds 
\begin{equation}
G_{\,\,t}^t=G_{\,\,r}^r=-2m\frac{\delta (r-\epsilon )}{r^2},  \label{previa}
\end{equation}
and 
\begin{equation}
G_{\,\,\theta }^\theta =G_{\,\,\phi }^\phi =-m\frac{\delta (r-\epsilon )}{r^2%
}-\frac m\epsilon \frac d{dr}\delta (r-\epsilon )=m\frac{\delta (r-\epsilon )%
}{r^2}-\frac{m\epsilon }{r^2}\frac d{dr}\delta (r-\epsilon ).
\label{previa1}
\end{equation}
We get in the limit $\epsilon \rightarrow 0$ 
\begin{equation}
G_{\,\,t}^t=G_{\,\,r}^r=-\frac 12G_{\,\,\theta }^\theta =-\frac 12%
G_{\,\,\phi }^\phi =-2m\frac{\delta (r)}{r^2},  \label{unexpected}
\end{equation}
which is exactly the result obtained in Ref.\cite{balasin93} using smoothed
versions of the Heaviside function $\Theta (r-\epsilon )$. This should be
contrasted with what is the expected result 
\begin{equation}
G_{\,\,b}^a=-8\pi m\delta ^3(\vec x)\delta _0^a\delta _b^0.  \label{expected}
\end{equation}
Note that the strategy followed here to obtain equations (\ref{previa},\ref
{previa1}) suggest the possibility of attain a distributional meaning to
curvature tensors by taking derivatives of the metric in the sense of
distributions, an idea we shall pursue  later. A second approach to the
problem \cite{balasin94}, using the Kerr-Schild {\it ansatz} for the metric
tensor, has been claimed to give the correct result (\ref{expected}),
assuming that even under the regularization procedure the metric maintains
its Kerr-Schild form. Nevertheless, it can be shown that in Kerr-Schild
coordinates, without the constraint imposed by the Kerr-Schild {\it ansatz},
the same $G_{\,\,b}^a$ as given by (\ref{unexpected}) is obtained.

Several regularization schemes of this kind can be questioned on formal
grounds. The unpleasant features are traceable, at least in part, to the
fact that the metric obtained from the line element (\ref{schcurv},\ref
{curvreg}) does not satisfies reasonable continuity requirements on the
3-surface $r=\epsilon $. Furthermore, as is well known for the Schwarzschild
solution in curvature coordinates, $r$ changes in character from a spacelike
coordinate for $r>2m$ to a timelike one for $r<2m$. Hence for values of $%
r<2m $ it is necessary to use a nonstatic system of coordinates. The
discontinuity of the metric tensor and the change of the signature cast
doubt over the previous results obtained from regularizing the Schwarzschild
metric in these coordinates.

In order to circumvent these problems we will consider a different approach
which is closer to the classical theory of distributions view and that has a
parallel into the theory of surface layers in General Relativity \cite
{israel}. Assuming that the Schwarzschild geometry can be considered as
being generated by an energy-momentum tensor with support in the
singularity, one method of attacking the problem is to replace the point
source by a uniform simple layer on a sphere of radius $\rho _0$. As $\rho
_0\rightarrow 0$ the configuration tends to a point but for radius $>\rho _0$
we always have an exact solution of Einstein's equations that looks exactly
like the Schwarzschild solution. In a certain sense, we shall have to deal
with a sequence of distributions, or with distributions depending on an
arbitrary parameter $\rho _0$.

The paper is organized as follows: in section {\bf II} we briefly recall the
construction of surface stress energy tensors for spherical thin shells,
following the treatment of junction conditions of Israel \cite{israel}. Some
features of this construction are also discussed. In section {\bf III},
using distributional techniques, the energy-momentum tensor on the whole
manifold for a spherical thin shell is derived and showed to be in complete
agreement with the results of previous section. Section {\bf IV} is devoted
to the point particle limit of the spherical shell, viewed as the limit of a
sequence of distributions. Following the same approach, it is shown in
section {\bf V}, how these results are extended to the Reissner-Nordstrom
spacetime.  Finally, some  concluding remarks are given in Section {\bf VI}. 

We shall use geometrized units in which $c=G=1,$ and the signature of the
metric is  $(-,+,+,+)$ 

\section{The Surface Layer Formalism.}

This section will be intented to apply the formalism of singular
hypersurfaces in order to construct the surface stress tensor of a spherical
thin shell. We shall follow Lake notation \cite{Lake}, with minor
differences in conventions.

Let us consider a spherical thin shell $\Sigma $, a singular time-like
hypersurface which divides the spacetime in two regions: the interior region 
$V^{-},$ described by flat minkowskian geometry and $V^{+},$ the exterior
spacetime, chosen as a spherically symmetric solution to Einstein field
equations. Both regions $V^{\pm }$ will be decribed by a single set of
isotropic coordinates, $x^a=(t,\rho ,\theta ,\phi )$ with respect to which
the line element is 
\begin{equation}
ds^2=-A^2dt^2+B^2(d\rho ^2+\rho ^2d\theta ^2+\rho ^2\sin ^2\theta \ d\phi ^2)
\label{elemento}
\end{equation}
where for the exterior geometry $A=A(\rho )$ and $B=B(\rho )$, are known
functions of the radial isotropic coordinate and for the interior region, $%
A=A(\rho _0)\equiv A_0$ and $B=B(\rho _0)\equiv B_0$ are constants. The
surface layer is represented in the chosen frame by 
\begin{equation}
f(\rho )=\rho -\rho _0=0  \label{layer}
\end{equation}
therefore, the components of the metric tensor matches naturally and
smoothly across the shell, corresponding to the well defined character of
the intrinsic geometry of $\Sigma ,$ in fact, the induced metric on $\Sigma $
takes the form 
\begin{equation}
ds_\Sigma ^2=-A_0^2dt^2+B_0^2\rho _0^2(d\theta ^2+\sin ^2\theta \ d\phi ^2)
\label{inducida}
\end{equation}
where intrinsic coordinates $\xi ^i=(t,\theta ,\phi )$ are used and the
3-metric elements induced are 
\begin{equation}
g_{ij}=g_{ab}\frac{\partial x^a}{\partial \xi ^i}\frac{\partial x^b}{%
\partial \xi ^j}  \label{ache}
\end{equation}
For future purpose we note here that the relation between coordinates of $%
V^{\pm }$ and intrinsic coordinates on $\Sigma $ are 
\begin{equation}
\frac{\partial x^a}{\partial \xi ^i}=\delta _{\,i}^a  \label{delta}
\end{equation}
On the other hand, the extrinsec geometry of $\Sigma $ is not well defined,
i.e., the extrinsec curvature tensor, ${\bf K\equiv -}\frac 12{\cal L}_{{\bf %
n}}{\bf g}$ induced on the layer is diferent as evaluated from its embedding
in $V^{+}$ or $V^{-}$ and the surface stress energy tensor is related to the
discontinuity of ${\bf K}$ through the Lanczos equation 
\begin{equation}
8\pi S_{\ j}^i=[K_{\ j}^i]-\delta _j^i\text{Tr}\left[ {\bf K}\right] \qquad
i,j=(t,\theta ,\phi )  \label{stress2}
\end{equation}
\[
S_{\ \,\rho }^\rho =S_{\ \,j}^\rho =0 
\]
where square brakets denotes a discontinuity across de layer as evaluated in 
$V^{+}$ and $V^{-}$, i.e., $\left[ f\right] \equiv f^{+}-f^{-}.$ Our task is
to perform the calculations to obtain the surface stress energy tensor of
the layer. From its definition it is easy to see that 
\begin{equation}
K_{ij}=-\nabla _bn_a\frac{\partial x^a}{\partial \xi ^i}\frac{\partial x^b}{%
\partial \xi ^j}  \label{K}
\end{equation}
$n_a$ being the components of ${\bf n,}$ the outward unit normal field to
the 3-surface $\Sigma $. Using (\ref{layer}) and (\ref{elemento}) the normal
is found to be 
\begin{equation}
n_a=B\delta _{\ a}^\rho  \label{normal}
\end{equation}
With the usual formula for the covariant derivative and equations (\ref
{delta}) and (\ref{normal}), equation (\ref{K}) yields the simple relation 
\begin{equation}
K_{ij}=\ B\,\Gamma _{\,\,ij}^\rho \ ,  \label{K2}
\end{equation}
where $\Gamma $ stand for the usual Christoffel symbols. Thus, raising
indexes, and performing the calculations for $V^{+}$ and for $V^{-}$ we
obtain 
\begin{equation}
\lbrack K_{\,t}^t]=-\frac{A^{\prime }}{AB},\qquad \qquad [K_{\,\,\theta
}^\theta ]=[K_{\,\,\phi }^\phi ]=-\frac{B^{\prime }}{B^2}  \label{K3}
\end{equation}
where and hereafter primes denotes differentiation with respect to $\rho $
and it is understood that quantities must be evaluated at $\Sigma $ after
performing the derivations. Finally, using equation (\ref{stress2}) for the
surface energy tensor, we arrive to the simple results 
\begin{equation}
S_{\,\,t}^t=\frac 1{4\pi B}\frac{B^{\prime }}B  \label{St}
\end{equation}
and 
\begin{equation}
S_{\,\theta }^\theta =S_{\,\,\phi }^\phi =\frac 1{8\pi B}\frac{(AB)^{\prime }%
}{AB}  \label{Steta}
\end{equation}

Let us apply the above results to obtain the surface stress tensor for a
spherical layer of total gravitational mass $m$. Therefore $A$ and $B$ will
be the metric elements corresponding to Schwarzschild vacumm solution in
isotropic coordinates 
\begin{equation}
A_S=(1-\frac m{2\rho })(1+\frac m{2\rho })^{-1}\qquad \text{in}\quad V^{+}
\label{Amas}
\end{equation}
and 
\begin{equation}
B_S=(1+\frac m{2\rho })^2\qquad \text{in\quad }V^{+}  \label{Bmas}
\end{equation}
where as usual, $m$ is the gravitational mass as measured at infinite. From (%
\ref{St}) and (\ref{Steta}), ${\bf S}$ can be readily obtained as 
\begin{equation}
S_{\,t}^t=\ -\frac m{4\pi \rho _0^2}(1+\frac m{2\rho _0})^{-3}  \label{St2}
\end{equation}
and 
\begin{equation}
S_{\;\theta }^\theta =S_{\,\,\phi }^\phi =\frac{m^2}{16\pi \rho _0^3}\ (1+%
\frac m{2\rho _0})^{-3}(1-\frac m{2\rho _0})^{-1}  \label{Steta2}
\end{equation}
The surface energy density of the layer $\sigma $, defined as $\sigma \equiv
S_{ij}u^iu^j$, where $u^i$ are the components of the intrinsic tangent
vector, $u^i=A^{-1}\delta _t^i,$ is given by $\sigma =$ $S_{\,t}^t.$ The
energy of the shell can be evaluated by integrating $\sigma $ over the
proper area of the $2$-sphere $t$= $const.$ and $\rho $ = $const.$ A simple
calculation gives 
\begin{equation}
{\cal E}=\int S_{ij}u^iu^j\ \,da=\int \sigma \,da=m+\frac{m^2}{2\rho _0}.
\label{energy}
\end{equation}

\section{The distributional approach to thin shells.}

Consider again the line element 
\begin{equation}
ds^2=-A^2(\rho )dt^2+B^2(\rho )(d\rho ^2+\rho ^2d{\theta }^2+\rho
^2sin\theta \,\,d{\phi }^2),  \label{schisot}
\end{equation}
\begin{equation}
A(\rho )=\cases{\displaystyle{\frac{(1 - \frac{m}{2\rho_0})}{(1 + 
\frac{m}{2\rho_0})}} &
$(\rho<\rho_0)$ \cr 
\displaystyle{\frac{(1 - \frac{m}{2\rho})}{(1 + \frac{m}{2\rho})}} &
$(\rho>\rho_0)$ \cr},  \label{schisot1}
\end{equation}
\begin{equation}
B(\rho )=\cases{(1 + \frac{m}{2\rho_0})^2 & $(\rho<\rho_0)$ \cr (1 +
\frac{m}{2\rho})^2 & $(\rho>\rho_0)$ \cr},  \label{schisot2}
\end{equation}
which corresponds to the geometry produced by a very thin spherical shell of
isotropic radius $\rho _0$: inside the sphere, the metric is equivalent to
the flat-space Minkowski metric and outside it agrees with the Schwarzschild
solution written in isotropic coordinates, provided we restrict ourselves to
values of $\rho _0$ satisfying the condition $\rho _0>\frac 12m$.

From (\ref{schisot}) we obtain for the Einstein tensor 
\begin{eqnarray}
G_{\,\,t}^t &=& \frac{2B^{\prime \prime }B\rho -\ B^{\prime 2}\rho
+4B^{\prime }B}{B^4\rho \ }\ ,  \label{einstein1} \\
\ G_{\,\,\rho }^\rho &=& \frac{B^{\prime 2}A\ \rho +2B^{\prime
}AB+2A^{\prime }B^{\prime }B\ \rho +2A^{\prime }B^2}{AB^4\rho },
\label{einstein2} \\
\ G_{\,\,\theta }^\theta &=&G_{\,\,\phi }^\phi = \frac{B^{\prime
}AB+B^{\prime \prime }AB\rho +A^{\prime }B^{\prime 2}-B^{\prime 2}A\rho
+A^{\prime \prime }B^2\rho }{AB^4\rho }  \label{einstein3}
\end{eqnarray}
Because of the linear character of the theory of distributions, the
differential equations satisfied by them should be linear. The above
expressions can not be written in the form of linear differential operators
acting on $A$ and $B$ nor even on some funtions of these. To what extent can
the Einstein tensor be regarded as a distribution? What we shall consider
here is an alternative that one should expect to find useful in many
nonlinear problems. In classical analysis, $A$ and $B$ are locally
integrable infinitely differentiable functions of variable $\rho $ except at 
$\rho =\rho _0$ where $A^{\prime }$ and $B^{\prime }$ are discontinuos and $%
A^{\prime \prime }$and $B^{\prime \prime }$ fail to exist. However,
differentiating $A$ and $B$ in the sense of distributions, $A^{\prime }$ and 
$B^{\prime }$ are now distributions with jump discontinuities at $\Sigma $
with resulting $f\delta (\rho -\rho _0)$ terms in the expressions for $%
A^{\prime \prime }$ and $B^{\prime \prime }$, with $f$ the corresponding
jump of the first derivative. From equations (\ref{einstein1}-\ref{einstein3}%
), this procedure provides us with a physically sensible well defined
distributional tensor $G_{\,\,b}^a$ given by 
\begin{eqnarray}
G_{\,\,t}^t &=&-\frac 1{\rho ^2AB^3}2m(1-\frac m{2\rho _0})\delta (\rho
-\rho _0),  \label{shell1} \\
G_{\,\,\rho }^\rho &=&0,  \label{shell2} \\
G_{\,\,\theta }^\theta &=&G_{\,\,\phi }^\phi =\frac 1{2\rho ^2AB^3}\frac{m^2%
}{\rho _0}\delta (\rho -\rho _0),  \label{shell3}
\end{eqnarray}
with support on the 3-surface $\Sigma $. This confirm the interpretation of
equations (\ref{schisot}-\ref{schisot2}) as the geometry whose source
corresponds to a uniform surface layer density spread on the 3-surface $%
\Sigma $, i.e., to an energy-momentum tensor ${\bf T}$ with support on $%
\Sigma $: 
\begin{eqnarray}
T_{\,\,\,t}^t &=&-\frac 1{8\pi \rho ^2AB^3}2m(1-\frac m{2\rho _0})\delta
(\rho -\rho _0),  \label{propiot} \\
T_{\,\,\,\rho }^\rho &=&0,  \label{propior} \\
T_{\,\,\,\theta }^\theta &=&T_{\,\,\,\phi }^\phi =\frac 1{16\pi \rho ^2AB^3}%
\frac{m^2}{\rho _0}\delta (\rho -\rho _0),  \label{propioth}
\end{eqnarray}
where use have been made of the Einstein equation ${\bf G}=8\pi {\bf T}$. We
recall that, the linear character of the theory of distributions avoids a
more rigorous justification of the above result. However, its correctness
follows directly from the relation between the surface energy-momentum
tensor $S_{\,\,\,b}^a$ on the 3-surface $\Sigma $, equations (\ref{St2},\ref
{Steta2}), and the energy-momentum tensor $T_{\,\,\,b}^a$ on the whole
manifold (\ref{propiot}-\ref{propioth}), namely \cite{misner} 
\begin{equation}
S_{\ \,b}^a\equiv \lim_{\epsilon \rightarrow 0}\Bigg[ \int_{-\epsilon
}^{+\epsilon }T_{\ \,b}^adn\Bigg],  \label{stress}
\end{equation}
with $dn=Bd\rho $ being the proper distance element measured perpendicularly
through the 3-surface $\Sigma $.

The energy-momentum tensor ${\bf T}$ can be cast in the form of the
energy-momentum tensor of an anisotropic fluid 
\begin{equation}
T_{\,\,b}^a=\zeta u^au_b+P_{\perp }h_{\,\,\,b}^a  \label{covariant}
\end{equation}
where $\zeta $ is the proper energy density, $u^a$ are the components of the
unit timelike four velocity, $P_{\perp }$ is the tangential pressure and $%
h_{\,\,\,b}^a$ is the projection tensor onto the subspace orthogonal to $%
{\bf u}$ and ${\bf n,}$ 
\begin{equation}
h_{\,\,\,b}^a=\delta _b^a+u^au_b-n^an_b  \label{proyector}
\end{equation}
with 
\begin{equation}
\zeta =\frac m{4\pi \rho ^2AB^3}(1-\frac m{2\rho _0})\delta (\rho -\rho _0)
\label{densidad de ener}
\end{equation}
and 
\begin{equation}
P_{\perp }=\frac{m^2}{16\pi \rho ^3AB^3}\delta (\rho -\rho _0).
\label{ptangencial}
\end{equation}
Note that the calculation of the energy of the layer as measured by an
observer with velocity ${\bf u}$ gives 
\begin{equation}
{\cal E}\equiv \int T_{ab}\,u^au^bd^3\!\sigma =\int \zeta \,d^3\!\sigma =m(1+%
\frac m{2\rho _0}),
\end{equation}
where $d^3\!\sigma $ is the volumen element of the rest space of the
observer, in complete agreement with (\ref{energy}). Furthermore, the Tolman
formula \cite{tolman,landau} for the total energy of a static and
asymptotically flat spacetime 
\begin{equation}
{\cal E}_T=\int (T_{\,\,\rho }^\rho +T_{\,\,\theta }^\theta +T_{\,\,\phi
}^\phi -T_{\,\,t}^t)\sqrt{-g}d^3x,  \label{tolman}
\end{equation}
with $g$ the determinant of the four dimensional metric and $d^3x$ the
coordinate volume element, gives 
\begin{equation}
{\cal E}_T=m,  \label{masa}
\end{equation}
as it should be. Interesting enough, note that the difference between the
total energy as measured in infinite and the energy of the shell, is $-%
m^2 /2\rho _0,$ which is the expression for the gravitational newtonian
energy of a thin shell; recall however that the radial coordinate $\rho _0$
is not the proper radius of the shell.

\section{The point particle limit and the Schwarzschild geometry.}

Now we consider the limit in which the spherical layer tends to a point
particle and let us comment on the limiting procedure. It should be noted
that the value of $ m/2\rho _0$ must be keep fixed and satisfying the
condition $ m/2\rho _0 <1$ in order to have a well defined metric
tensor, equations (\ref{schisot},\ref{schisot1},\ref{schisot2}) for all $%
\rho $. Furthermore, the static character of this spacetime reduce the
distributional evaluation of $\lim_{\rho _0\rightarrow 0}G_{\,\,b}^a$ to a
three-dimensional problem on the 3-surfaces $t=const.$ For $ m/2\rho _0
$ fixed, it turns out that the sequence of distributions $G_{\,\,b}^a$ is
such that $\lim_{\rho _0\rightarrow 0}\left\langle G_{\,\,b}^a,\Phi
\right\rangle $ exist for each $\Phi $. Let us ilustrate the evaluation of $%
\lim_{\rho _0\rightarrow 0}G_{\,\,b}^a$ with the explicit calculation of $%
\lim_{\rho _0\rightarrow 0}G_{\,\,t}^t$ where the condition $ m/2\rho _0%
= fixed$ is understood. We have 
\begin{eqnarray*}
\left\langle G_{\,\,t}^t,\Phi (\vec x)\right\rangle _{\vec x}
&=&\left\langle -\frac 1{\rho ^2AB^3}2m(1-\frac m{2\rho _0})\delta (\rho
-\rho _0),\Phi (\vec x)\right\rangle _{\rho ,\theta ,\phi } \\
\ &=&-\int_0^\infty d\rho \int_0^\pi d\theta \int_0^{2\pi }d\phi \,\rho
^2B^3\sin \theta \,\,\,\frac 1{\rho ^2AB^3}2m(1-\frac m{2\rho _0})\delta
(\rho -\rho _0)\Phi (\rho ,\theta ,\phi ) \\
\ &=&-2m(1+\frac m{2\rho _0})\int_0^\pi d\theta \int_0^{2\pi }d\phi \sin
\theta \,\,\Phi (\rho _0,\theta ,\phi ).
\end{eqnarray*}
Taking into account that 
\[
\lim\limits_{\rho _o\rightarrow 0}m(1+\frac m{2\rho _0})=\lim\limits_{\rho
_o\rightarrow 0}\left( m+2\rho _0(\frac m{2\rho _0})^2\right) =m, 
\]
it follows that 
\begin{eqnarray*}
\lim\limits_{\rho _o\rightarrow 0}\left\langle G_{\,\,t}^t,\Phi (\vec x%
)\right\rangle _{\vec x} &=&\lim\limits_{\rho _o\rightarrow 0}-2m(1+\frac m{%
2\rho _0})\int_0^\pi d\theta \int_0^{2\pi }d\phi \sin \theta \,\,[\Phi (\vec 
0)+(\Phi (\rho ,\theta ,\phi )-\Phi (\vec 0)] \\
&=&\lim\limits_{\rho _o\rightarrow 0}\left[ -8\pi m(1+\frac m{2\rho _0})\Phi
(\vec 0)-2m(1+\frac m{2\rho _0})\int_0^\pi d\theta \int_0^{2\pi }d\phi \sin
\theta \,\,(\Phi (\rho ,\theta ,\phi )-\Phi (\vec 0)\right] \, \\
&=&-8\pi m\Phi (\vec 0)=\left\langle -8\pi m\delta ^3(\vec x),\Phi (\vec x%
)\right\rangle _{\vec x}.
\end{eqnarray*}
Hence 
\begin{equation}
\lim\limits_{\rho _o\rightarrow 0}G_{\,\,t}^t=-8\pi m\delta ^3(\vec x)
\label{getete}
\end{equation}
The evaluation of the limit for the remaining non zero components of $%
G_{\,\,b}^a$ proceeds in the same manner. The results are 
\begin{equation}
\lim\limits_{\rho _o\rightarrow 0}G_{\,\,\theta }^\theta =\lim\limits_{\rho
_o\rightarrow 0}G_{\,\,\phi }^\phi =0  \label{gfifi}
\end{equation}
It them follows that this limit is a well defined distribution given by 
\begin{equation}
G_{\,\,\,b}^a=-8\pi m\delta ^3(\vec x)\delta _0^a\delta _b^0,
\end{equation}
which implies {\it via} Einstein equations, 
\begin{equation}
T_{\,\,\,b}^a=-m\delta ^3(\vec x)\delta _0^a\delta _b^0,  \label{final}
\end{equation}
as expected.

\section{The Reissner-Nordstrom Spacetime.}

Next, let us consider the case in which the shell has a uniformly
distributed charge $e$. Hence, the $V^{+}$ spacetime line element is given
by the Reissner-Nordstrom exterior solution to the Einstein-Maxwell
equations, appropiately written in isotropic coordinates, i.e., 
\begin{equation}
A_{RN}=\frac{(1-\displaystyle{\frac{m^2}{4\rho ^2}})+
\frac{e^2}{4\rho ^2}}{(1+\displaystyle{\frac m{%
2\rho }})^2-\frac{e^2}{4\rho ^2}} \qquad \text{in}\quad V^{+}
\label{amenos}
\end{equation}
and 
\begin{equation}
B_{RN}=(1+\frac m{2\rho })^2-\frac{e^2}{4\rho ^2}\qquad \text{in\quad }V^{+}
\label{bmenos}
\end{equation}
It follows from (\ref{St}) and (\ref{Steta}) that 
\begin{equation}
S_{\,\,\,t}^t=\ -\frac 1{8\pi \rho _0^2}\left[ 2m(1+\frac m{2\rho _0})-\frac{%
e^2}{\rho _0}\right] B_{RN}^{-2}(\rho _0)  \label{St3}
\end{equation}
and 
\begin{equation}
S_{\;\,\theta }^\theta =S_{\,\,\,\phi }^\phi =\frac{m^2-e^2}{16\pi \rho _0^3}%
\ B_{RN}^{-2}(\rho _0)\ A_{RN}^{-1}(\rho _0).  \label{Steta3}
\end{equation}
Some comments are in order. First observe that in the extreme
Reissner-Nordstrom solution, $e=\pm \,m,$ the tangential stresses vanish.
This is explained if we realize that $e=\pm \,m$ is precisely the Bonnor
condition \cite{bonnor} for a selfgravitating pressureless charged fluid to
be in equilibrium, therefore in this case we are dealing with a dust thin
shell. Secondly, note that the calculation of the energy contained in the
charged layer, integrating (\ref{St3}) as was done in (\ref{energy}), gives 
\begin{equation}
{\cal E}=m(1+\frac m{2\rho _0})-\frac{e^2}{2\rho _0},  \label{energia2}
\end{equation}
the last term in this equation representing the (negative) electrostatic
contribution to the energy. Finally, in the limit $e=0$ these equations
reduce to the neutral case, as expected.

Following the same approach of section {\bf III}, the energy-momentum tensor
for the spherical thin shell with a uniformly distributed charge $e$ can be
obtained. It is given by 
\begin{eqnarray}
T_{\,\,t}^t &=&-\frac 1{8\pi \rho ^2A_{RN}B_{RN}^3}\Bigg[\frac{(1-\frac{m^2}{%
4\rho _0^2})+\frac{e^2}{4\rho _0^2}}{(1+\frac m{2\rho _0})^2-\frac{e^2}{%
4\rho _0^2}}\Bigg] \bigg[2m\big(1+\frac m{2\rho _0}\big) -\frac{e^2}{\rho _0}%
\bigg] \delta (\rho -\rho _0)  \nonumber \\
&&-\frac{e^2}{8\pi \rho ^4B_{RN}^4}\Theta (\rho -\rho _0),  \label{RNt} \\
T_{\,\,\rho }^\rho &=&-\frac{e^2}{8\pi \rho ^4B_{RN}^4}\Theta (\rho -\rho
_0),  \label{RNr} \\
T_{\,\,\theta }^\theta &=&T_{\,\,\phi }^\phi =\frac 1{16\pi \rho
^2A_{RN}B_{RN}^3}\frac{m^2-e^2}{\rho _0}\delta (\rho -\rho _0)  \nonumber \\
&&\quad \quad +\frac{e^2}{8\pi \rho ^4B_{RN}^4}\Theta (\rho -\rho _0).
\label{RNth}
\end{eqnarray}
From these results one can verify, using equation (\ref{stress}), that the
energy-momentum tensor ${\bf S}$ on the hipersurface $\Sigma $, equations (%
\ref{St3},\ref{Steta3}) is in fact the ''delta-like singularity'' at $\Sigma 
$ \cite{misner} of the energy-momentum tensor ${\bf T}$ given by (\ref{RNt}-%
\ref{RNth}). Note that the non-delta-like electromagnetic part of (\ref{RNt}-%
\ref{RNth}) is traceless. Furthermore, from (\ref{tolman}) and (\ref{RNt}-%
\ref{RNth}), it follows the remarkable result ${\cal E}_T=m$, i.e., ${\cal %
E_T}$ no contains electromagnetic contributions.

The limiting case of a point particle runs over the same lines given in
Section {\bf IV}. The result is 
\begin{eqnarray}
T_{\,\,t}^t &=&-m\delta ^3(\vec x)-\frac{e^2}{8\pi \rho ^4B_{RN}^4}\Theta
(\rho ),  \nonumber \\
T_{\,\,\rho }^\rho &=&-\frac{e^2}{8\pi \rho ^4B_{RN}^4}\Theta (\rho ), 
\nonumber \\
T_{\,\,\theta }^\theta &=&T_{\,\,\phi }^\phi =\frac{e^2}{8\pi \rho ^4B_{RN}^4%
}\Theta (\rho ).
\end{eqnarray}
as expected.

\section{Conclusions and remarks.}

We have shown that a succesfull approach for dealing with curvature tensor
valued distribution is to first impose admisible continuity conditions on
the metric tensor, and then take its derivatives in the sense of classical
distributions. The distributional meaning is then equivalent to the junction
condition formalism. Afterwards, through appropiate limiting procedures, it
is then possible to obtain well behaved distributional tensors with support
on submanifolds of $d\leq 3$, as we have shown for the energy-momentum
tensors associated with the Schwarzschild and Reissner-Nordstrom spacetimes.
The above procedure provides us with what is expected on physical grounds.
However, it should be mentioned that the use of Colombeau's new generalized
functions \cite{colombeau} in order to obtain distributional curvatures \cite
{clarke,steinbauer}, may renders a more rigorous setting for discussing
situations like the ones considered in this paper.

\section*{Acknowledgments}

We kindly thank Luis Herrera and Umberto Percoco for fruitful discussions
and constant interest in this work.

\end{document}